\documentstyle[eqsecnum,aps,prb,multicol]{revtex}

\begin{document}
\draft

\title{Surface wave scattering at nonuniform fluid
interfaces}

\author{T. Chou and David R. Nelson}

\address{Dept. of Physics, Harvard University,
Cambridge, MA 02138}

\date{\today}
\maketitle

\begin{abstract}
Effects of spatially varying interfacial parameters on the
propagation of surface waves are studied.  These
variations can arise from inhomogeneities in coverage of
surface active substances such as amphiphillic molecules
at the fluid/gas interface. Such variations often occur in
phase coexistence regions of Langmuir monolayers.  Wave
scattering from these surface inhomogeneities are
calculated in the limit of small variations in the surface
parameters by using the asymptotic form of surface
Green's functions in the first order Born approximation.
When viscosity and variations in surface elastic moduli
become important, modes other than transverse capillary
waves can change the characteristics of propagation.
Scattering among these modes provides a mechanism for
surface wave attenuation in addition to viscous damping
on a homogeneous surfactant covered interface.
Experimental detection of waves attenuation and
scattering is also discussed.
\end{abstract}

\vspace{5mm}

\pacs{PACS numbers: 68.35.Ja, 47.35.+i, 92.10.H,
68.10.-m}

\begin{multicols}{2}
\narrowtext

\section{Introduction}

Investigation of interfacial phenomena has attracted
much attention, especially if interfacial properties are
modified by a surface active material. Many studies
involve amphiphillic molecules at the fluid/air interface
with monolayer or less coverage.  Small amounts of
material at an interface can drastically alter wave
propagation and also lead to complex  phase behavior.
Recent experiments with fluorescence and Brewster angle
microscopies have sought to elucidate the structure of
monolayer films of fatty acids, cholesterol, and
phospholipids.\cite{REV} A myriad of phase transitions of
the surface film are directly observable as well as
theoretically predicted.\cite{THS}

These studies often reveal high surfactant concentration
domains of various shapes and sizes in the coexistence
regions.  Figures 1(a) and 1(b) show Brewster angle
microscope images of circular and striped domains of
triglycerides in the Liquid Condensed/Liquid Expanded
and Liquid Expanded/Gaseous ``coexistence regions''
respectively.  The figures span about 1mm. Lighter
regions correspond to regions of higher surfactant
concentrations.  This type of coverage is often associated
with a phase coexistence region not unlike that of the
gas-liquid phase transition.  These first-order like
transitions can be described approximately by a
thermodynamic Maxwell construction across
mechanically unstable regions of the equation of state.
External electric fields can also be used to generate local
interfacial surfactant concentration
gradients.\cite{AY,KM} If nonlocal interactions such as
dipole forces are important, regular modulated phases
appear in the usual two-phase coexistence region.
\cite{THS}

Experiments have also investigated changes in capillary
wave propagation caused by insoluble surfactants at an
air- water surface.\cite{EXP,TC1}  Wave damping
experiments on film covered air/water interfaces show
a remarkable increase in the damping rate above that
caused by viscous bulk fluid dissipation.  Even a single
monolayer of surface active substances, such as
amphiphillic fatty acids, increases the damping rate by
as much as five fold by altering the flow trajectories
beneath the surface.  In addition, at monolayer
concentrations, the damping coefficients appear to be
sensitive to nonuniform surface coverages.  At
concentrations where an apparent coexistence of
surfactant phases occurs, the damping rate peaked at
five or six times that due to viscous
dissipation.\cite{TC1}

Attempts to explain the propagation of water waves
covered with surfactant have considered the
homogeneous mechanical properties of floating films
\cite{DOR}.  Although the theories of
Dorrenstein,\cite{DOR} Levich,\cite{LEV}, and
Luccassen\cite{LUC,HA} all predict a maximum in the
damping factor, they are incomplete since they rely on
surface parameters which are not well defined in a
nonuniform surface.  Though a peak in damping rate is
experimentally observed, these systems are almost always
inhomogeneous.  As argued in Refs. (11) and (12), wave
propagation should depend on the heterogeneous nature
of the interface.  The details of this surface
inhomogeneity depend on how the film was applied as
well as the phase diagrams of these surface materials.

Wave scattering off of surface heterogeneities can serve
as an additional damping or attenuation mechanism.
Much as light is attenuated in a turbid medium, surface
waves can be scattered by inhomogeneities like those in
Fig. 1.  Gravity wave scattering from a discontinuity in
the trough depth has been calculated\cite{MEI} and
extended to multiple scattering and wave localization in
a random bottom by Belzons.\cite{LOC}  Single scattering
of gravity waves off obstacles or depth discontinuities
has been well studied \cite{MEI} and has been applied to
multiple scattering of gravity waves off obstacles
protruding from the water surface\cite{YUE}.  Work done
in this area has usually been in the context of
oceanography and has focussed on gravity waves
propagating in the presence spatial variations of the
boundaries, particularly the finite depth bottom.

We will consider capillary waves scattering off variations
in parameters such as surface elastic constants and
surface tension. An exact solution of water wave
scattering from discontinuities in boundary condition
parameters can be found for very special cases \cite{CS};
in this paper we will use approximation methods which
are valid for small variations in surface tension and
interfacial modulus.

In the next Section,  we  present a model for an inviscid,
infinite depth substrate fluid with a spatially varying
surface tension.  Regions of sharply different surface
tensions are stabilized by line tensions at their
boundaries.  This line tension could be supplied, for
example, by a waxed circular thread floating at the
interface and separating a surfactant rich monolayer
from a clean fluid interface. This model problem leads
naturally to a Green's function describing the response
due to a point normal force acting on a uniform surface.

In Section III the asymptotic limit of this Green's
function is used in a Born approximation to describe
the scattering of potential flow by a circular domain of
different surface tension.  The scattering function is
used to calculate the attenuation of a plane capillary
wave via a scattering cross section.

The effects of bulk viscosity and surface viscoelasticity
are included in Section IV.  Variations in surface {\it
compressibility} are the primary source of scattering in
this context since the surface {\it tensions} of coexisting
island phases like those in Fig. 1 are nearly equal.  Here
the tangential stresses mix both potential and vorticity
fields.  The lowest order Green's function matrix is
derived.  The analysis shows the existence of at least
three distinct surface modes: transverse, ``in-plane''
longitudinal, and ``in-plane'' shear.

Scattering via the Born approximation for a viscoelastic
surface is calculated in Section V using the Green's
functions of Section IV.  The dissipation of energy into
the various channels is also considered. Wave
amplitude and energy attenuation due to these
processes is discussed in Section VI, where
experimental implications are also considered.

\section{Model for Surface Tension Variation on an
Inviscid Fluid}

In this section we investigate capillary wave
propagation in the presence of surface tension
inhomogeneities with an inviscid Newtonian fluid
occupying the half space $z\leq 0.$ This first model is
somewhat artificial because if one hypothesizes a
surface tension discontinuity due to two distinct
surfactant phases at an interface, one phase would
immediately grow at the expense of the other to reduce
the free energy. With short ranged interactions only, the
surface tensions of two phases coexisting in
equilibrium are necessarily equal. The surface tension
changes across modulated phases with large scale
structures determined by dipole-dipole interactions
are also small, vanishing as the reciprocal island
dimension for the configurations shown in Fig. 1.
Allowing for fixed surface tension variations is
nevertheless instructive and might even be realizable
experimentally over a reasonable time period by
confining one phase of an insoluble monolayer inside a
flexible barrier of negligible stiffness.

The mechanical stress balance at the interface in the
presence of a spatially varying surface tension
$\sigma(x)$ reads,\cite{LANDAU}

\begin{equation}
{\bf \hat{n}}\cdot({\bf \Pi}_{+} -  {\bf \Pi}_{-}) =
\sigma(\vec{r}) \left({1\over R_{1}} + {1\over
R_{2}}\right){\bf
\hat{n}} - \vec{\nabla}_{\!s}\,\sigma(\vec{r}),
\label{BC}
\end{equation}

\noindent where ${\bf \hat{n}}$ is the unit vector normal
to the interface, $\vec{\nabla}_{\!s}$ is a tangential
gradient on the interface, and $R_{1}$ and $R_{2}$ are
the local principle radii of curvature of the interface.
${\bf \Pi}_{+}$ and ${\bf \Pi}_{-}$ denote the stress
tensor above and below the interface respectively.  On an
inhomogeneous surface, ${\bf \Pi}$ and $\sigma$ may
depend on the surface coordinates; the projection of
these coordinates onto the $x-y$ plane is denoted by
$\vec{r}$. For ideal fluids, the stress tensor is
proportional to the pressure.  In the Monge
representation, the linearized normal component of
Eq. (\ref{BC}) is,

\begin{equation}
\pi(z=0) =\sigma(\vec{r}) \nabla^{2}_{\!\perp}
\eta(\vec{r}), \label{BCN}
\end{equation}

\noindent where $\eta$ is the vertical displacement of
the interface from its equilibrium position at $z=0$, and
$\vec{\nabla}_{\!\perp}$ is a gradient in the $x-y$
plane. Eq. (\ref{BCN}) is the usual dynamic boundary
condition which determines the dispersion relation for
surface waves.  Here, $\pi$ is the jump in dynamic
pressure across the interface. We assume that the
tangential stresses are balanced by surface forces such
as line tension.

For an ideal fluid, the motion will be irrotational and the
velocity field can be expressed via a potential,

\begin{equation}
v_{i}= \partial_{i} \phi (\vec{r},z)
\end{equation}

\noindent satisfying $\nabla^{2} \phi(\vec{r}, z) =0$
in the bulk fluid. At the surface we have,

\begin{equation}
\partial_{t} \eta(\vec{r}) \simeq
\partial_{z}\phi(\vec{r}, z=0),
\end{equation}

\noindent and the Navier-Stokes equations lead to,

\begin{equation}
\pi(\vec{r},z) = -\rho \partial_{t} \phi(\vec{r},z) -
g\eta(\vec{r}), \label{P}
\end{equation}

\noindent where $\pi(\vec{r},z)$ is the fluid
pressure, $\rho$ is the fluid density and $g$
the gravitational constant.  Henceforth, gravity will be
neglected ($g=0$). The equation for the velocity
potential at the surface reads,

\begin{equation}
\left[ -\partial_{t}^{2} +
(\sigma_{0}+\delta\sigma(\vec{r}))\nabla_{\perp}^{2}
\partial_{z}\right]\phi(\vec{r}, z)\,
\rule[-3mm]{.2mm}{8mm}\,_{z=0}=0
\end{equation}

We now introduce a harmonic time dependence
$\omega$ to all quantities and set $\rho \equiv 1$.
Upon adding a unit $z-$component delta function force
to Eq. (\ref{BCN}), we are led to define a surface Green's
function which solves

\begin{equation}
\left[ \omega^{2} +(\sigma_{o}+\delta \sigma(\vec{r}))
\nabla^{2}_{\perp} \partial_{z} \right] G(\vec{r},0)
= -i\omega \delta(\vec{r}).\label{BCN2}
\end{equation}

\noindent  The deviation of the spatially varying
surface tension from the constant value $\sigma_{0}$
is denoted by $\delta \sigma(\vec{r})$ . Equation
(\ref{BCN2}) can be solved for $\delta \sigma \equiv
0$ by expanding $G_{0}$ in a general Fourier
integral,

\begin{equation}
G_{0}(\vec{r}, z, \omega) = \int {d^{2}k
\over (2\pi)^{2}}\,G_{0}(k)
e^{i \vec{k}\cdot \vec{r}+\vert k \vert z},
\end{equation}

\noindent The solution to (\ref{BCN2}) can then be
formally written as,

\begin{equation}
G(\vec{k})= G_{0}(k) + G_{0}(k)\int{d^{2}q\over (2\pi)^{2}}\,
q^{2}\vert q \vert\,
\delta\sigma(\vec{k}-\vec{q})\,G(\vec{q})\label{GI}
\end{equation}

\noindent with

\begin{equation}
\sigma_{0} G_{0}(k) = (k^{2}\vert k \vert
-p^{3})^{-1}\label{G0k}
\end{equation}

\noindent where $p \equiv (\omega^{2} /
\sigma_{0})^{1/3}$ is real and positive. This transverse
mode described by $G_{0}(k)$, $i.e.,$ capillary waves
propagate at $k=\pm p.$

\section{Scattering from a single circular domain}

In this section, the scattering of a plane inviscid capillary
wave from a single circular domain with a sharp surface
tension discontinuity is calculated. The surface tension
inside a circular domain of radius $a$ is $\sigma_{1} =
\sigma_{0} + \delta \sigma$.  A perimeter whose only
effect is to confine the domain and maintain the
discontinuity is assumed. A schematic of this scattering
problem is depicted in Figure 2. The wedge at the left is
typically an electrocapillary wavemaker, excited at
frequency $\omega$. \cite{SMK}

Two approximations will be made. First, a transverse
capillary wave impinging on a domain will change its
perimeter; we neglect line tension energy of the
confining loop in comparison to the areal stretching
energy of the surface.  Second, we will assume a frozen
variation in surface tension due to a fixed domain.  This
last simplification requires weak scattering, {\it i.e.,} $ka
\ll 1$, where $k$ is the wavevector and $a$ is the
domain radius.

For a single domain, a two dimensional analog of the
Young-Laplace equation can be defined for circular
domain of radius $a$ with a sharp surface tension
discontinuity.

\begin{equation}
\sigma_{0}-\sigma_{1} = {\gamma \over 2 a}\label{YL}
\end{equation}

\noindent where $\sigma_{0}$ and $ \sigma_{1}$ are
the surface tensions inside and outside the domain
respectively and $\gamma$ is the line tension acting
through the circular domain boundary. Equation
(\ref{YL}) is actually the tangential component of
(\ref{BC}) when bulk viscosity is neglected. For small
$\delta\sigma$, the scattering is obtained in the first
iteration or Born approximation.  In the inviscid case, the
elastic scattering mechanism described here will be the
only process resulting in attenuation of an incident plane
wave in the midst of a collection of scattering centers.

By using Eq. (\ref{G0k}) , we can find the exact surface
Green's function, $G_{0}(\vec{r})$,  for a homogeneous
interface,

\begin{eqnarray}
G_{0}(r) = {i\omega\over
\sigma_{0}}\int{d^{2}k\over (2\pi)^{2}}\, {e^{-i
\vec{k}\cdot \vec{r}}\over k^{2}\vert k \vert
-p_{+}^{3}}\;\;\;\;\;\;\;\;\;\;\;\;\;\;\;\;\;\;\; \nonumber \\
\;\;\;= -{\omega\over 6
\sigma_{0}}H_{0}^{(1)}(pr) + {i\omega\over \pi^{2}
\sigma_{0}p}\int_{0}^{\infty} \,{K_{0}(yr)y^{4} \over y^{6}
+ p^{6}}dy \label{G02}
\end{eqnarray}

\noindent In the above integrands, $p_{+} \equiv
p+i\epsilon$ represents a small damping which
removes the ambiguity in the integrals for outgoing
Green's functions.  Since $p$ is real, (\ref{G02}) is
dominated by the first term at large distances ($pr \gg
1$) and the asymptotic form becomes,

\begin{equation}
G_{0}(r) \sim {\omega \over
3\sigma_{0}p}{e^{ipr+3\pi i/4} \over \sqrt{2 \pi p r}}
\end{equation}

\noindent With the help of the Green's function,
$G_{0}(r),$ the solution of  Eq. (\ref{BCN2})  can be
written in integral form as,

\begin{equation}
\phi(r,0) = e^{ipx} + {i\over \omega} \int d^{2}r'\,
\delta\sigma(r')\,G_{0}(\vert \vec{r}-\vec{r}\,' \vert)
\nabla\,'^{2}\partial_{z} \phi(r', 0)
\label{I0}
\end{equation}

The first term of (\ref{I0}) represents an incoming
plane wave with unit magnitude at the origin of the
scatterer generated by a $z$ component force at
$x\rightarrow -\infty$  such that the power law
evanescent waves represented by the second term in
the last line of (\ref{G02}) have long died off.  The
integral term is the scattered wave due to the
``potential'' $\delta\sigma(r')$. In the lowest order
approximation for small $\delta\sigma(r'), \phi(r')$ in
the integral is replaced with $\phi_{inc}(r') =
e^{ipx'}$. In the far field, the potential takes the form,

\begin{equation}
\phi(r) \equiv e^{ipx} + f(\theta){e^{ipr} \over \sqrt{r}}
\label{ASYM1}
\end{equation}

\noindent where the scattering factor is the asymptotic
limit of the integral in (\ref{I0}),

\begin{equation}
f(\theta) = -i\sqrt{2\pi i  p}\,{a \over 3}\left({\delta\sigma
\over \sigma_{0}}\right){J_{1}(pa sin {\theta \over 2})
\over sin {\theta \over 2}}\label{f}
\end{equation}

\noindent Validity of this approximation is ensured
when the scattered wave just outside the disc is small
compared to the incident plane wave, which leads to
the requirement,

\begin{equation}
(pa)^{3}\left({\delta\sigma \over \sigma_{0}} \right)^{2}
\ll 1.
\end{equation}

\noindent In this asymptotic limit, the total scattering
length is,

\begin{equation}
\Sigma_{s} = \int_{0}^{2 \pi} d\theta \vert
f(\theta)\vert^{2} \,\simeq \,  {\pi^{2} a \over
9}\left({\delta\sigma \over \sigma_{0}} \right)^{2}
(pa)^{3}.
\label{SIGMA}
\end{equation}

\noindent Upon using the capillary wave dispersion
formula, $\omega = \sqrt{\sigma_{0}p^{3}}$, we see
that Eq.  (\ref{SIGMA}) implies that $\Sigma_{s} \sim
\omega^{2}$, in contrast to the famous Rayleigh
result $\Sigma_{s} \sim \omega^{4}$ for light
scattering from the atmosphere \cite{JACKSON}.

Nonuniform coverage of air/liquid interfaces
however, usually results in a collection of circular
scatterers.  Scattering from an uncorrelated collection
of identical domains (assumed for simplicity here to
be of identical size) reduces the forward coherent
intensity according to,

\begin{equation}
\vert \langle \phi(x) \rangle \vert ^{2} \propto e^{-n
\Sigma_{s}(p) x} \label{BEER}
\end{equation}

\noindent where $n$ is the number of scattering
domains per area.  Beer's law, equation (\ref{BEER}),
applies in the single scattering regime and is valid
when the typical inter-domain distance $l \gg 2\pi/p
\gg a$; our calculation for $\Sigma_{s} \ll a$ only
describes weak attenuation accurately. An alternative
calculation which averages over surface disorder and
leads to the same result is given in Appendix B.

\section{Model With Bulk Viscosity and Surface
Elasticity}

In actual experimental systems such as the one
depicted in Figure 2, surface layers have additional
parameters which are spatially varying.  For example,
the surface compressibility will be different inside
and outside the islands even if the surface tension is
uniform.  The restoring forces proportional to these
surface parameters are coupled to the bulk viscous
forces at the interface.  Surface elasticity is an
important factor in the propagation and attenuation
of surface waves \cite{LEV} and can arise from
surfactant molecules at the interface.  Fluid motion in
the interfacial plane tends to locally distort the film
which resists due to its surface compressive or shear
elasticity and interfacial 2D viscosity.

The normal component of (\ref{BC}) now reads,

\begin{equation}
(\sigma_{0} + \delta
\sigma(\vec{r}))\nabla_{\perp}^{2} v_{z} -i\omega \pi
(\vec{r}, z) + 2i\omega \nu \partial_{z} v_{z}\,
\rule[-3mm]{.2mm}{8mm}\,_{\stackrel{ }{z=0}}  =
-i\omega f_{z}(\vec{r}) \label{N1}
\end{equation}

\noindent where $\sigma_{0}$ is the uniform
background surface tension, $\delta \sigma(\vec{r})$
is the spatial variation in surface tension considered
earlier, $\nu$ is the bulk fluid viscosity, and
$f_{z}(\vec{r})$ is proportional to an external pressure in
the $z$-direction.

Upon considering the intrinsic stresses of the
interface, we find that the {\it tangential} forces  are
balanced when,

\begin{equation}
\vec{\nabla}_{\!\perp}\sigma(r) +
\nabla_{\!\perp}\!\cdot{\bf \Pi}^{'(2)} -\nu (\partial_{z}
\vec{v}_{\perp} + \vec{\nabla}_{\!\perp} v_{z})\,
\rule[-3mm]{.2mm}{8mm}\,_{\stackrel{ }{z=0}}=
\vec{f}_{\perp}(\vec{r}), \label{T1}
\end{equation}

\noindent where $\vec{f}_{\perp}(r)$ is proportional to an
in-plane stress applied at the interface. The first term on
the left side of (\ref{T1}) arises from gradients in surface
tension and acts like a pressure term in the stress tensor
of an ideal two dimensional liquid.  The dissipative part of
the 2D fluid stress tensor is given by ${\bf \Pi}^{'(2)}$. In
the remainder of the paper, we assume
$\vec{\nabla}_{\!\perp}\!\cdot{\bf \Pi}^{'(2)} = 0$; all of
the phenomena can be adequately accounted for by
$\vec{\nabla}_{\!\perp}\sigma(\vec{r})$. Modifications
due to dissipative and elastic contributions to ${\bf
\Pi}^{'(2)}$ is considered in Appendix A.

Fluid motion at the surface convects surface molecules
and changes the surfactant concentration
$\Gamma(\vec{r},t)$  locally, so we set

\begin{equation}
\vec{\nabla}_{\!\perp}\sigma(r) = {d \sigma \over d
\Gamma} \vec{\nabla}_{\!\perp} \Gamma(\vec{r}).
\label{D1}
\end{equation}

\noindent For an insoluble material, conservation of
surfactant at the interface gives,\cite{LEV}

\begin{equation}
\partial_{t}\Gamma(\vec{r}) + \vec{\nabla}_{\!\perp}
\!\cdot (\vec{v}_{\perp} \Gamma(\vec{r})) \simeq
-i\omega\delta \Gamma(\vec{r}) + \Gamma_{0}
\vec{\nabla}_{\!\perp} \!\cdot \!\vec{v}_{\perp} = 0,
\label{CON}
\end{equation}

\noindent with $\Gamma(\vec{r}) \equiv \Gamma_{0}
+ \delta \Gamma(\vec{r})$, where $\Gamma_{0}$ is
the equilibrium surfactant concentration inside or
outside the domain. Equations (\ref{D1}) and
(\ref{CON}) can be combined to express the surface
tension gradient in phase $j$ in terms of the
in-plane velocity field,

\begin{equation}
\vec{\nabla}_{\!\perp} \sigma(\vec{r}) \simeq {i\over
\omega} B_{j}\vec{\nabla}_{\!\perp}
(\vec{\nabla}_{\!\perp}\!\cdot\vec{v}_{\perp})\, \,
\rule[-3mm]{.2mm}{8mm}\,_{\stackrel{ }{z=0}},
\label{DS}
\end{equation}

\noindent where the bulk modulus is defined by

\begin{equation}
B_{j} \equiv -\Gamma_{j} {d\sigma \over d\Gamma}
\rule[-3mm]{.2mm}{8mm}\,_{\stackrel{ }{\Gamma =
\Gamma_{j}}} \label{B}
\end{equation}

\noindent with $j=0,1$ representing $B$ outside or
inside a domain for example. We can now solve
equations (\ref{N1}) and (\ref{T1}). Upon
decomposing the velocity into its potential and
vorticity components,

\begin{equation}
v_{i}=\partial_{i} \phi + \epsilon_{ijk} \partial_{j} \psi_{k}
\end{equation}

\noindent we choose a gauge where $\psi_{z} = 0$,
and define $ \beta_{i} \equiv \epsilon_{ij}\psi_{j}$
(the vorticity is $\vec{\Omega} = \vec{\nabla} \times
\vec{v} = -\hat{z}\times\nabla^{2}\vec{\beta}$) so
that,

\begin{equation}
\vec{v}=\vec{\nabla} \phi(r) + \left( \begin{array}{c}
\partial_{z} \vec{\beta} \\ -\vec{\nabla}_{\!\perp}\!\cdot
\vec{\beta} \end{array} \right),
\end{equation}

\noindent where the vector $\vec{\beta}$ lies in the
$x-y$ plane.  The fields $\phi$ and $\vec{\beta}$
satisfy,

\begin{equation}
\nabla^{2} \phi(\vec{r}) = 0 \label{POT}
\end{equation}

\noindent and

\begin{equation}
\partial_{t} \vec{\beta} = \nu \nabla^{2}
\vec{\beta}\label{DIFF}
\end{equation}

\noindent respectively, which are solved by  the
expansions,

\begin{equation}
\phi(\vec{r}, z) = \int{d^{2}k\over
(2\pi)^{2}}\,\phi(\vec{k})\,e^{i \vec{k} \cdot \vec{r}+ \vert k
\vert z} \label{Pk}
\end{equation}

\noindent and

\begin{equation}
\vec{\beta}(\vec{r}, z) = \int{d^{2}k\over (2\pi)^{2}}\,
\vec{\beta}(\vec{k})\, e^{i \vec{k} \cdot \vec{r} +
lz}\label{Bk}
\end{equation}

\noindent where  $l \equiv \sqrt[+]{k^2-i\omega/
\nu}$.  The positive roots for $l$ and $\vert k \vert $
are taken to ensure that $\phi$ and $\beta$ vanish as
$z \rightarrow -\infty$.

Finally, upon using (\ref{Pk}),  (\ref{Bk}),  (\ref{DS}),
(\ref{P}) in (\ref{N1}), and taking a time derivative of
(\ref{T1}), the normal and tangential stress balances at the
surface $z \simeq 0$ after Fourier transforming in the
in-plane $\vec{r}$ coordinate take the form,

\begin{equation}
{\bf L}(q) \mbox{\boldmath $\psi$}(\vec{q}) =
\mbox{{\it \boldmath $F$}}(\vec{q}) + \int{d^{2}k\over (2\pi)^{2}}\, {\bf
M}(\vec{k},\vec{q})\, \mbox{\boldmath $\psi$} (\vec{k})
\label{IE}
\end{equation}

\noindent where \mbox{{\it \boldmath $ F$}}$(\vec{q})$
is the $q^{th}$ Fourier component of $(f_{z},f_{x},f_{y})$.
In (\ref{IE}), $\mbox{\boldmath $ \psi$}(\vec{r}) \equiv (
\phi, \beta_{x}, \beta_{y})$ and $\vec{k}$ and $\vec{q}$
are wavevectors in the $x-y$ plane.  Spatial variations in
the surface tension $\sigma(\vec{r})$ and interfacial bulk
modulus $B$ are included in the matrix
\end{multicols}
\widetext

\begin{eqnarray}
 {\bf M}(\vec{k},\vec{q}) \equiv {i \over \omega}\left[
\begin{array}{ccc} \delta\sigma(\vec{q}-\vec{k})k^{2}\vert k
\vert & \delta\sigma(\vec{q}-\vec{k})k^{2}k_{x} &
\delta\sigma(\vec{q}-\vec{k})k^{2}k_{y} \\
\delta B(\vec{q}-\vec{k}) i k^{2}k_{x} & \delta
B(\vec{q}-\vec{k})k_{x}^{2}l & \delta
B(\vec{q}-\vec{k})k_{x}k_{y}l \\
\delta B(\vec{q}-\vec{k})ik^{2}k_{y} & \delta
B(\vec{q}-\vec{k})k_{x}k_{y}l & \delta B(\vec{q}-\vec{k})
k_{y}^{2}l \end{array} \right].
\end{eqnarray}

\noindent When the film is homogeneous, ${\bf M} = {\bf
0}$.

The coefficients $\sigma$ and $B$ in ${\bf L}(\vec{q})$
are constants, {\it e.g.} the values $\sigma_{0}$ and
$B_{0}$ outside an isolated circular domain, or mean
values appropriate for a heterogeneous collection of
many domains.  In equilibrium, for large domains, the
surface tension contrast will be small as discussed above.
Hence, in the following analysis (unlike that for the
nonviscous fluid), we make the simplifying assumption
that the surface tension is constant in monolayers in
coexistence and near equilibrium.  Scattering will be
solely due to the contrast in $B$. The compressional
modulus for phase $j=(0,1)$, $B_{j}$, can be measured by
the slopes of the isotherms just outside the coexistence
plateau as shown by the dotted lines in Fig. 3.  These
values will be different in the two phases even for
monolayers in equilibrium.

\noindent The matrix ${\bf L}(\vec{q})$ corresponding to
an interface with $B=B_{0}$ is,

\begin{eqnarray}
{\bf L}(\vec{q}) \equiv \left[ \begin{array}{ccc} i\left[
\omega-{\sigma_{0} q^{2}\over \omega}\vert q \vert
+2i\nu q^{2}\right]  &  q_{x}\left[2i \nu l -{\sigma_{0}
q^{2}\over \omega}\right] & q_{y}\left[2i\nu
l-{\sigma_{0}
q^{2}\over \omega} \right]\\
-iq_{x}\left[{iB_{0} \over \omega} q^{2}+2\nu \vert q
\vert\right] & -\left[ {iB_{0}\over \omega}
q_{x}^{2}l+\nu(l^2+q_{x}^{2})\right] &
-q_{x}q_{y}\left[ {iB_{0} \over \omega}l+\nu\right] \\
-iq_{y}\left[{iB_{0} \over \omega} q^{2}+2\nu \vert q
\vert\right] & -q_{x}q_{y}\left[ {iB_{0} \over \omega} l
+\nu\right] & -\left[ {iB_{0}\over \omega}
q_{y}^{2}l +\nu(l^2+q_{y}^{2})\right] \end{array}
\right]
\end{eqnarray}

\begin{multicols}{2}
\narrowtext

\noindent  Upon inverting (\ref{IE}), the equation for
\mbox{\boldmath $\psi$} takes the form

\begin{equation}
\mbox{\boldmath $\psi$}(\vec{q}) = {\bf G}(q) \mbox{{\it
\boldmath $F$}} (\vec{q}) + {\bf G}(q) \int{d^{2}k\over
(2\pi)^{2}}\,{\bf M}(\vec{k}, \vec{q})\,\mbox{\boldmath
$\psi$}(\vec{k}), \label{IE2}
\end{equation}

\noindent with a matrix Green's function ${\bf G}(\vec{q})
\equiv {\bf L^{-1}}(\vec{q}).$  For fixed $\omega$, the
poles of the Green's function give the wavevectors of the
possible modes. For the region of parameter space most
accessible to experiments, there are two poles
corresponding to outgoing capillary waves. Appendix A
shows that if $\nabla_{\!\perp}\!\cdot{\bf \Pi}^{'(2)} \neq
0$, additional modes corresponding to in-plane motion
also exist.

When viscosity is small, the ``transverse'' capillary wave
pole is near the inviscid pole; it is shifted from $k=p$ by a
small real part and develops a small imaginary part due to
viscous damping.  When $x \equiv \nu \left( \omega/
\sigma_{0}^{2} \right)^{1/3} \ll 1$,

\begin{equation}
k_{t} \simeq p+ \Delta= p+\Delta'+i\Delta''\label{KT}
\end{equation}

\noindent The functions, \cite{LEV,HA}

\begin{equation}
\Delta' \simeq {1 \over 3}\left( {\omega^{2} \over
\sigma_{0}} \right)^{1/3} {y^{2}\sqrt{x/2}-yx \over
y^{2}-y\sqrt{2x}+x} \label{DELTA1}
\end{equation}

\noindent and,

\begin{equation}
\Delta'' \simeq {1 \over 3}\left( {\omega^{2} \over
\sigma_{0}} \right)^{1/3} {y^{2}\sqrt{x/2}-2yx\sqrt{2x}
+4x^{2} \over y^{2}-y\sqrt{2x}+x}\label{DELTA2}
\end{equation}

\vspace{2mm}

\noindent  are plotted in Figure 4(a) as a function of $y
\equiv B_{0}/\sigma_{0}$.  For $\omega \ll
B^{2}/\nu^{3}$ the longitudinal pole is approximately,

\begin{equation}
k_{l} \simeq {\nu^{1/4} \omega^{3/4} \over
\sqrt{B_{0}}}e^{i\pi /8}+\ldots \label{KL}
\end{equation}

\noindent The relative dissipation between the
transverse and longitudinal modes,
$Im\{k_{t}\}/Im\{k_{l}\}$, are plotted in Figure 4(b). The
``shear'' mode found in the Appendix A can usually be
omitted  since it has a large imaginary part and is highly
damped.  The asymptotic form of the Green's function
including the transverse and longitudinal modes is
explicitly displayed in Appendix C.

\section{First order scattering from viscoelastic domains}

To consider the effects of surface elasticity, the inclusion
of a nonzero bulk viscosity is necessary to satisfy the
tangential force balance equation. Thus, there is inelastic
wave attenuation even without surface inhomogeneities
and elastic wave scattering.  In this Section, we calculate
the scattering of surface waves from domains of different
modulus $B$ only. The variation in the modulus, $\delta
B = B_{1}-B_{0}$, of domains near equilibrium can be
estimated from the isotherm construction in Fig. 3.  Since
the periods of the surface waves are probably smaller
than the molecular diffusion times, the relevant quantity
is the isoentropic modulus rather than the isothermal
modulus found from Fig. 3. We assume equality of surface
tensions inside and outside the domains.

Incident waves with various amounts of transverse and
longitudinal (and possibly shear) mode components can
be generated with surface forces at various angles from
the normal. The incoming wave we consider will be that
generated by a force which acts solely in the $z$
direction at the surface. The relative amounts of
transverse and longitudinal components is fixed by the
coefficients of the relevant Green's functions given in
Appendix C.

In the large $r$ limit, Eq. (\ref{IE2})  yields the form,

\begin{eqnarray}
\mbox{\boldmath $\psi$}(\vec{r}) \simeq
\mbox{\boldmath $\psi$}_{0}(x) +
\sum_{a,b=t,l}\vec{g}_{a,b}(\theta){e^{ik_{a}r} \over
\sqrt{r}} \label{ASYM2}
\end{eqnarray}

\noindent where \(\mbox{ \boldmath $\psi$}_{0}(x)\) is a
plane normalized combination of transverse and
longitudinal waves generated by a force per length
$f_{z}$ with infinite extent in the $y$ direction. The wave
induced by the wavemaker will initially have transverse
and longitudinal modes with both $\phi$ and
$\beta_{x}$ components. By integrating the asymptotic
forms of the Green's function from Appendix C with
respect to the direction along the wavemaker blade, we
can determine \(\mbox{\boldmath $\psi$}_{0}\) to
leading order in $\nu$.  The vertical interfacial
displacement generated by a plane wavemaker,

\begin{equation}
\eta(x) \simeq \left[{-i\over
3\sigma_{0}p} e^{ik_{t}x} + {e^{5\pi i/8}\nu^{5/4} \over
2B_{0}^{3/2}\omega^{1/4}}e^{ik_{l}x}\right] f_{z},
\end{equation}

\noindent determines the nature of the waves incident
on the scattering domains. If $Im\{k_{t}\} \ll  Im\{k_{l}\},$
incident waves are mainly transverse capillary waves;  if
$Im\{k_{t}\} \gg Im\{k_{l}\},$ the incident waves are
predominately  longitudinal waves.  The ratio of the
imaginary parts of the transverse and longitudinal
modes are plotted in Fig. 4(b) for various values of $B$.

The form factor in Eq.  (\ref{ASYM2}) describes a
complicated scattering process.  The $\phi$ and
$\vec{\beta}$ components of $\mbox{\boldmath
$\psi$}$ scatter among themselves; as suggested by the
indexes $a,b = t,l$ on the scattering amplitudes
$g_{a,b}(\theta).$ The components of
$\vec{g}_{a,b}(\theta)$ governing the scattering of $\phi$
has four terms describing scattering between transverse
and longitudinal modes. To lowest order in $\omega
\nu^{3} / \sigma_{0}^{2}$ and $\omega \nu^{3}
/B_{0}^{2}$ these are,

\begin{equation}
g_{t,t}(\theta) \simeq {4a^{2}\sqrt{\pi \nu} p^{7/4} \over
\sigma_{0}^{1/4}}e^{i\pi /4} \left({\delta B \over
B_{0}}\right){J_{1}(2ap\,sin \frac{\theta}{2}) \over
2ap sin\frac{\theta}{2}} \,cos \theta, \label{ftt}
\end{equation}

\begin{eqnarray}
g_{t,l}(\theta) \simeq e^{5\pi i /8}a^{2} \sqrt{\pi}
p^{17/8}{(2B_{0}-\sigma_{0})\nu^{5/4} \over B_{0}^{3/2}
\sigma_{0}^{1/8}} \left({\delta B \over
B_{0}}\right)\nonumber \\
\times {J_{1}(qa) \over qa}\,cos \theta,
\label{ftl}
\end{eqnarray}

\begin{eqnarray}
g_{l,l}(\theta) \simeq e^{i\pi/4} \sqrt{2\pi
k_{l}}\,  (pa)^{2}
{3\sigma_{0}(2B_{0}-\sigma_{0})\nu^{2}\over
4B_{0}^{3}}\left({\delta B \over B_{0}}\right)\nonumber \\
\times{J_{1}(2ak_{l}sin{\theta \over 2}) \over
2ak_{l}sin{\theta \over 2}}\,cos \theta\label{fll}
\end{eqnarray}

\noindent and,

\begin{equation}
g_{l,t}(\theta) \simeq a^{2}p^{9/4}\sqrt{{\pi \over
2}}{\sigma_{0}^{5/4} \nu^{3/2}\over B_{0}^{2}}\left(
{\delta B \over B_{0}} \right) {J_{1}(qa)\over qa}\,cos
\theta.\label{flt}
\end{equation}

\noindent In the above equations, $q \equiv
\sqrt{k_{t}^{2}+k_{l}^{2} -2k_{t}k_{l}cos \theta}.$ Each
of these scattering functions correspond to a
particular intermode scattering.  For example,
$g_{t,l}(\theta)$ represents the scattering of an
impinging transverse wave into a longitudinal wave.
The two limiting cases depicted in Figures 5(a) and
5(b) will be of interest.

In Fig. 5(a), the longitudinal mode is more dissipative
than the transverse mode. For modes described by Eqns.
(\ref{KT}), (\ref{DELTA1}), (\ref{DELTA2}), and (\ref{KL}),
this first case corresponds to $Im\{k_{l}\} \gg l^{-1} \gg
\Delta''$. In this case, we may assume that only
transverse capillary waves reach the scattering centers.
Some of the transverse amplitude, however, is scattered
into the longitudinal mode which is then quickly
viscously dissipated leading to enhanced inelastic
attenuation of the velocity potential. For an areal density
of identical island scatterers $n$, the transmitted
potential decays according to, \cite{AI}

\begin{equation}
\langle \vert \phi(x) \vert \rangle\sim exp[-{n\over
2}(\Sigma_{t,t}+\Sigma_{t,l})x-\alpha x] \label{POTD}
\end{equation}

\noindent where the scattering and absorption lengths,
$ \Sigma_{t,t}$ and $\Sigma_{t,l}$ are given by angular
integrals over $g_{t,t}(\theta)$ and $g_{t,l}(\theta)$,

\begin{equation}
\Sigma_{t, t} \simeq
{4(pa)^{4} \pi^{2} \nu \over \sqrt{\sigma_{0}p}}
\left({\delta B \over B_{0}} \right)^{2} \label{Stt}
\end{equation}

\noindent and,

\begin{equation}
\Sigma_{t,l} \simeq  {(pa)^{4}\pi^{2}\nu^{5/2}
p^{1/4}\over 4 B_{0}^{3}\sigma_{0}^{1/4}}
(2B_{0}-\sigma_{0})^{2} \left( {\delta B \over
B_{0}}\right)^{2}. \label{Stl}
\end{equation}

\noindent To make these and subsequent expressions
dimensionally meaningful, let $\sigma_{0}, B_{0}
\rightarrow \sigma_{0}/\rho, B_{0}/\rho$ where $\rho =
1$ g/cm$^{2}$ for the interface between air and water.
Since $\omega \sim p^{3/2}$ for capillary waves, these
scattering lengths depend on frequency and viscosity
according to $\Sigma_{t,t} \sim \nu\omega^{7/3}$, and
$\Sigma_{t,l} \sim \nu^{5/2}\omega^{17/6}.$ Note that
$\Sigma_{t,t}$ describes an {\it elastic} scattering
process, which could be detected by placing a wave
detector at an angle to the incident wavevector, and
measuring the scattering amplitude $g_{t,t}(\theta)$
directly.  $\Sigma_{t,l}$ acts as an enhanced viscous
dissipation under the conditions described above. In our
approximations, $\Sigma_{t,t} \gg \Sigma_{t,l}$; the
major scattering process is elastic. Only when
$\sigma_{0}/B_{0}$ is very large can $\Sigma_{t,l}$ be
comparable to $\Sigma_{t,t}$. The damping from viscous
dissipation is described by $\alpha = \Delta''$ displayed
in Eq.  (\ref{DELTA1}).

In Fig. 5 (b), the opposite situation obtains, $\Delta''
>Im\{k_{l}\}$. The predominant mechanism is
scattering of longitudinal waves. Here, the relevant
scattering lengths are,

\begin{equation}
\Sigma_{l,l} \simeq
9\pi^{2}(pa)^{4}\,{\sigma_{0}^{19/8}p^{9/8}\nu^{17/4}
(2B_{0}-\sigma_{0})^{2} \over 32 B_{0}^{13/2}}
\left({\delta B \over B_{0}}\right)^{2} \label{Sll}
\end{equation}

\noindent and,

\begin{equation}
\Sigma_{l,t} \simeq
\pi^{2}(pa)^{4}\,{\sigma_{0}^{5/2}\nu^{3}\sqrt{p} \over
8B_{0}^{4}} \left({\delta B \over B_{0}}\right)^{2},
\label{Slt}
\end{equation}

\noindent which behave as $\Sigma_{l,l} \sim
\nu^{17/4}\omega^{41/12}$ and $\Sigma_{l,t} \sim (\nu
\omega)^{3}$. The approximations
$\omega\nu^{3}/B_{0}^{2} \ll 1$ and
$\omega\nu^{3}/\sigma_{0}^{2} \ll 1$ give $\Sigma_{l,t}
\gg \Sigma_{l,l}$; in this case, the predominant
scattering is that of longitudinal waves scattering into
transverse waves which are then quickly dissipated by
the bulk viscosity.  Again, only when  $\sigma_{0}/B_{0}
\gg 1$ is the elastic longitudinal $\rightarrow$
longitudinal scattering comparable.  Here, simple viscous
attenuation of longitudinal waves behaves as,

\begin{equation}
\alpha = Im\{k_{l}\}={\sqrt{2-\sqrt{2}}\over 2}\,\,
{\nu^{1/4}\omega^{3/4} \over \sqrt{B_{0}}}.
\end{equation}

\section{Implications for Experiments}

Surface wave damping is expected to have
experimental signatures with the techniques used in
recent studies.  Induced surface waves can be probed
by light scattering and detecting the beam deflection
or measuring the power spectrum. \cite{TC1}
Alternatively, direct mechanical measurements due to
the force from impinging waves are also used.

Our calculations for one scatterer can be tested by
scattering an induced wave off a single isolated domain.
The asymptotic radial distribution of scattered potential
should follow those given in equations (\ref{ASYM1}) and
(\ref{ASYM2}) for the elastic (inviscid)  and inelastic
(viscous) cases respectively.  Because a single surfactant
domain is rarely seen, an isolated domain may have to be
constructed by placing a thin confining thread on the
surface of a homogeneous monolayer and aspirating the
material outside of this loop.  Provided the solubility into
the bulk liquid is small, this arrangement, although
metastable, may be quite long-lived.  In practice, the
scattering from the thread loop alone may have to be
considered.  Electric fields have also been used to
manipulate surfactant concentrations and orientations at
the interface.\cite{AY,KM}  Single circular domains of
surfactant rich or depleted phases have been produced
with this method.\cite{KM}

Damping of induced height fluctuations propagating
through an array of scattering domains can be directly
measured by probing the amplitude as a function of
distance from the source. Experimentally, the nucleation
of surfactants results in domains of various sizes and
shapes.  Our analysis describes a monodisperse
collection of scatterers but can be extended to include
polydispersity by averaging over domain size
distributions.

Using Eqs. (\ref{SIGMA}) and (\ref{BEER}) as guides, an
estimate of the scattering effect can made. For a
membrane which has only a surface tension variation,
scattering attenuation above that of viscous damping
will be noticeable when,

\begin{equation}
n \left(\delta\sigma
\over \sigma_{0}\right)^{2}\geq
{ \nu \omega\over a(pa)^{3}\sigma_{0}}
\label{DAMP}
\end{equation}

\noindent The value $\delta\sigma$ is difficult to
measure directly.  The surface tension difference inside
and outside a domain of radius $a$ is related through Eq.
(\ref{YL}); however, these quantities are difficult to
measure but would be expected to be small for systems
near equilibrium. The plateaus in the isotherms are rarely
perfectly flat due to long-ranged dipolar interactions
between surface molecules and the slow relaxation to
equilibrium.  An experimental estimate\cite{MG} gives
line tensions on the order of $\gamma \simeq 10^{-6}$
erg/cm. For domains of radius $\sim 0.1$ mm, this
tension gives $\delta\sigma \simeq 10^{-3}$ dyne/cm.
For a fluid with $\sigma_{0} \simeq 70$ dyne/cm driven
at a frequency $\omega  = 2\pi\times1000$ s$^{-1}$
and covered with $\sim 10^{3}$ scatterers per cm$^{2}$,
Eq.  (\ref{DAMP}) requires unphysical bulk viscosities
$\nu \leq 10^{-13}$ cm$^{2}$/s before attenuation due
to scattering will be noticeable.

When viscosity is explicitly included, the transverse wave
attenuation given by (\ref{POTD}) includes a damping
term $-\Delta''x$ which has a maximum (Fig.  4(a)) as
predicted by homogeneous theories of Dorrenstein and
others.\cite{DOR,LEV}  For the coexistence of fatty acids
in the low coverage regime, the modulus $B$ is expected
to be much smaller that the surface tension; thus is not
clear whether the maximum seen in experiments can be
described  by this peak.  However, for wavelengths much
larger than the typical interdomain distance, a definition
of an effective $B$ may be consistent with observations.
Appendix A shows that if only surface viscosities
$\eta^{(2)}$ and $\zeta^{(2)}$ are important, $\Delta''$
has no predicted peak.

The attenuation effects of elastic and inelastic scattering
are measured by the scattering lengths (\ref{Stt}),
(\ref{Stl}), (\ref{Sll}), and (\ref{Slt}), as well as the viscous
dissipation coefficients $\Delta''$ and $Im\{k_{l}\}$.
Detection of interfacial scattering via damping is feasible
when the relevant $n\Sigma$ is comparable to $\Delta''$
or $Im\{k_{l}\}$. In the transverse $\rightarrow$
transverse wave scattering, this condition leads to,

\begin{equation}
n\left( {\delta B\over B_{0}}\right)^{2} \geq
{p^{7/4}\sigma_{0}^{1/4} \over \pi^{2}(pa)^{4}
\sqrt{\nu}}. \label{stt}
\end{equation}

\noindent For $\omega = 2\pi\times1000$ s$^{-1}$,
$\sigma_{0} \sim 70$ dyne/cm, $\nu = 0.01$
cm$^{2}$/s, and $a \sim 1\,$mm, Eq. (\ref{stt}) yields
$n(\delta B /B_{0})^{2} \geq 1.4$. Note that this
experimentally realizable criterion for detecting
scattering through wave attenuation depends strongly on
$pa$, elastic scattering having a larger effect when the
wavelength is smaller than the domain radius $a$.  For
the process involving longitudinal wave scattering
pictured in Fig. 5(b), the attenuation is proportional to
$n\Sigma_{l,t}$. The criterion for observing inelastic
scattering assisted damping is,

\begin{equation}
n\left({\delta B \over B_{0}}\right)^{2} \geq
{p^{5/8} B_{0}^{7/2} \over
\nu^{11/4}\sigma_{0}^{17/8} \pi^{2} (pa)^{4}}.
\label{slt}
\end{equation}

\noindent Though it is possible to find parameters such
that Eqs. (\ref{stt}) and (\ref{slt}) are satisfied, the
accuracy of the Born approximation is questionable
when $pa$ or $(\delta B/B_{0})$ are too large.  However,
experiments do show enhanced damping when the
transverse capillary wavelength is small compared to the
typical domain size, $(pa > 1)$,   and no observable
damping above viscous dissipation for large
wavelengths. \cite{WFM}

Alternatively, the scattered waves may be directly
detected.  The incident waves vectors are scattered in
magnitude (in the inelastic case) and direction. Using
light scattering, this experiment may be realized by
rotating the light scattering plane an angle $\theta$ away
from the direction of the incident wavevector as shown in
Figure 6. For a small number of elastic scatterers ($\nu
=0$), the signal would be proportional to $\vert f(\theta)
\vert ^{2}$ where the scattering function is given by Eq.
(\ref{f}). At higher densities, the random scattering is
expected to effectively contribute to the thermal ripplons
near the source. A difference in amplitude of the thermal
spectra peak measured at $\theta = \pi/2$ is expected
depending on whether the wave maker is on.

\acknowledgements
T.C. wishes to acknowledge helpful discussions with
Vassilios Houdzoumis.  We have also benefitted from
discussions with Howard Stone.  The authors wish to
thank Qing Yu Wang and Aryeh Feder for Figure 1.
Support from the National Science Foundation through
the Harvard Materials Research Laboratory and through
grant DMR-91-15491 is gratefully acknowledged.

\appendix
\section{Intrinsic Surface stress effects}

When surface viscosities are included in the dissipative
part of the two dimensional stress tensor ${\bf
\Pi}^{'(2)}$, the behavior of the surface modes can
change. For a two dimensional Newtonian liquid,

\begin{equation}
\Pi^{'(2)}_{ij} = \eta^{(2)}
(\partial_{i}v_{j}+\partial_{j}v_{i}-\delta _{ij}
\partial_{k} v_{k})+\zeta^{(2)}\delta_{ij}\partial_{k}v_{k},
\label{PI2}
\end{equation}

\noindent where $\eta^{(2)}$ and $\zeta^{(2)}$ are the
two dimensional shear and dilatational viscosities
respectively. Using (\ref{PI2}) in (\ref{T1}), the tangential
stress balance becomes,

\begin{eqnarray}
\nabla_{\!\perp}\sigma(\vec{r}) +
\eta^{(2)}\nabla_{\perp}^{2}\vec{v}_{\perp}+\zeta^{(2)}
\vec{\nabla}_ {\!\perp}(\vec{\nabla}_{\!\perp}\!\cdot
\vec{v}_{\perp})\;\;\;\;\nonumber \\
\;\;\;\;\;-\nu(\partial_{z}\vec{v}_{\perp}+
\vec{\nabla}_{\!\perp}v_{z}) =
\vec{f}_{\perp}(\vec{r}).
\end{eqnarray}

In Section IV, we neglected surface viscosities. If however,
$B/\omega \ll \eta^{(2)}, \zeta^{(2)}$, then the transverse
pole is given by (\ref{KT}) with

\begin{equation}
\Delta' \simeq {\sqrt{2}\over 6}\left({\omega^{2}\over
\sigma_{0}}\right)^{1/3}{x^{3/2}y(y+4) \over
y^{2}x+y\sqrt{2x}+1}
\end{equation}

\noindent and,

\begin{equation}
\Delta'' \simeq {1\over 3}\left({\nu\omega\over
\sigma_{0}}\right) {y^{2}\sqrt{x/ 2} +y +4 \over
y^{2}x+y\sqrt{2x}+1}
\end{equation}

\noindent In the above expression, $x \ll 1$, and $y
\equiv (\eta^{(2)}+\zeta^{(2)})\omega^{2/3} / \nu
\sigma^{1/3}$.  Both $\Delta'$ and $\Delta''$ are
monotonic in $y$. The longitudinal mode for $\omega
\gg \nu^{3}/(\eta^{(2)}+\zeta^{(2)})^{2}$ becomes,

\begin{equation}
k_{l} \simeq {(\omega \nu)^{1/4}\,e^{3\pi i/8} \over
\sqrt{\eta^{(2)} +\zeta^{(2)}}} + \ldots \label{KLF}
\end{equation}

\noindent and a ``shear'' mode appears at,

\begin{eqnarray}
k_{s} = {\nu \over
\sqrt{2}\eta^{(2)}}\sqrt{1-\sqrt{1-{4i\omega\eta^{(2) 2}
\over \nu^{3}}}} \;\;\;\;\;\;\;\;\nonumber \\
\;\;\;\;\;\;\;\;\simeq \sqrt{{i\omega \over
\nu}}\left[1-{i\omega\eta^{(2) 2} \over 2\nu^{3}} +
\ldots\right] \label{KSF}
\end{eqnarray}

\noindent where the last approximation assumes
$\omega \ll \nu^{3}/\eta^{(2) 2}.$

We can also consider the case of a purely elastic interface;
a relevant example may be a polymerized membrane on
the air/water interface. For an isotropic elastic
membrane, $\nabla_{\!\perp}\!\cdot {\bf \Pi}^{'(2)}$ is
replaced with $-\delta F / \delta\vec{u}$ with the free
energy of the elastic sheet,

\begin{equation}
F={1 \over 2} \int d^{2}r \left[ \lambda(r)
u_{ii}^{2}+2\mu(r) u_{ij}^{2} \right] \label{F}
\end{equation}

where $u_{ij}= {1 \over
2}(\partial_{i}u_{j}+\partial_{j}u_{i})$ is a
two-dimensional strain tensor and $u_{i}$ are the
in-plane displacements from equilibrium.  $\lambda$
and $\mu$ are spatially varying  Lam\'{e} coefficients
which may be $\lambda_{0}, \mu_{0}$ outside the
domains and $\lambda_{1}, \mu_{1}$ inside.  With this
model, the transverse and longitudinal wavevectors are
given by (\ref{KT}), (\ref{DELTA1}), (\ref{DELTA2}), and
(\ref{KL}) respectively, but with an effective interfacial
modulus given by

\begin{equation}
B_{0}= 2\mu_{0} + \lambda_{0}-\Gamma_{0}{d\sigma \over
d\Gamma}\rule[-3.5mm]{.2mm}{9mm}\,
_{\stackrel{}{\Gamma = \Gamma_{0}}} .
\end{equation}

\noindent For $\mu \neq 0$, an elastic shear  mode also
occurs at,

\begin{equation}
k_{s}= {i \over \sqrt{2}}({\omega \nu \over
\mu})\sqrt{1-\sqrt{1+{4i \mu^{2} \over \omega \nu^{3}}}}
\simeq \sqrt{{i \omega \over \nu}}\left[ 1-i{\mu^{2}
\over 2 \nu^{3} \omega} + \ldots \right]\label{KSS}
\end{equation}

\noindent where the last approximation is valid when
$\omega \gg \mu^{2}/ \nu^{3}$.  The modes given by
(\ref{KT}), (\ref{DELTA1}), (\ref{DELTA2}), (\ref{KL}), and
(\ref{KSS}) describe a solid with vanishing shear modulus
or a liquid membrane with small surface viscosities.
Surface viscoelasticity of the interface can be
incorporated by allowing $B$ to be complex:
$B\rightarrow B+i\omega C.$

\section{Disorder Averaged Green's Function $\langle
G \rangle$}

In this section, we develop another means of calculating
the attenuation of the coherent velocity potential in the
viscid model. In the limit of small variations in the
surface tension and sparse uncorrelated scatterers, the
result is identical with that obtained from the first Born
iteration.  Equation (\ref{GI}) can be represented
diagrammatically as in Figure 7(a).  The result of
ensemble averaging over uncorrelated domains, $\langle
\delta\sigma(r) \delta \sigma(r') \rangle \simeq
\epsilon\,\delta(r-r')$, is shown in Figure 7(b).

In this summation, the cross terms representing
correlated scattering are neglected. Inverting the series
and defining an ensemble averaged Green's function,

\begin{equation}
\langle G(k)\rangle  \simeq G_{0}(k) \sum_{n=0}^{\infty}
\left[ \epsilon\, G_{0}(q)k^{3} \int^{\Lambda}_{0}
{d^{2}q\over (2\pi)^{2}}\,q^{2}\vert q
\vert G_{0}(q)\right] ^{n}\label{A1}
\end{equation}

The cut-off $\Lambda^{-1}$ gives a measure of the
typical spacing between domains of different surface
tension. This equation is an expansion in the small
quantity $\epsilon$. After performing the integral over
$G_{0}(q)$ and formally summing the series, the
propagator can written in the form,

\begin{equation}
\langle G(k)\rangle^{-1} =
{\sigma_{R}-i\sigma_{I}\over \sigma_{0}} \vert k \vert
k^{2}-p^{3}\label{A2}
\end{equation}

The form of $\langle G(k) \rangle^{-1}$ is that of
$G_{0}(k)^{-1}$, but with $\sigma_{0}$ replaced by an
effective complex surface tension where
$\sigma_{I}=\epsilon p^{2} /6 \sigma_{0}$ and
$\sigma_{R}$ is a $\Lambda$ dependent real part
proportional to $\epsilon$.  The Fourier transform of
$G(k)$ yields,

\begin{equation}
\langle G(k) \rangle \simeq {-ie^{2i\theta} \over 6p
\left[\sigma_{0}(\sigma_{R}^{2}+\sigma_{I}^2)\right]^{1/
3}}H_{0}^{(1)}(k_{0}r) + \ldots \label{A3}
\end{equation}

\noindent where $\theta \equiv \frac{1}{3}
tan^{-1}(\sigma_{I}/\sigma_{R})$ and,

\begin{equation}
k_{0} = p\left( {\sigma_{0} \over
\sqrt{\sigma_{R}^{2}+\sigma_{I}^{2}}}
\right)^{1/3}e^{i\theta}\label{A4}
\end{equation}

\noindent The attenuation of a one dimensional plane
wave is found by integrating the $y$
coordinate in (\ref{A3}),

\begin{equation}
\int_{-\infty}^{\infty} \langle G(r) \rangle dy \propto
e^{ik_{0}x} \sim e^{-\alpha x}.
\end{equation}

\noindent where the damping factor $\alpha$ is,

\begin{eqnarray}
\alpha = p\left( {\sigma_{0} \over
\sqrt{\sigma_{I}^{2}+\sigma_{R}^{2}}}
\right)^{1/3}\!\!\!sin\left[
\frac{1}{3}tan^{-1}(\sigma_{I}/\sigma_{R})\right]\;\;
\nonumber \\
\;\;\;\;\;\;\; \simeq
{\epsilon\, p^{3} \over 18\sigma_{0}^{2}} +
O(\epsilon^{2}) \label{ALPHA}
\end{eqnarray}

The correlation $\epsilon$ for randomly distributed,
noninteracting domains  can be found by considering
the averages,

\begin{equation}
\langle \sigma(\vec{r})\rangle = \sigma_{0} c +
\sigma_{1}(1-c)
\end{equation}

\noindent and,

\begin{equation}
\langle \sigma(\vec{r})^{2} \rangle =
\sigma_{0}^{2}c+\sigma_{1}^{2}(1-c) - \langle
\sigma(\vec{r}) \rangle^{2},
\end{equation}

\noindent where $c$ is the area fraction occupied by
the domains. Hence,

\begin{equation}
\langle\delta\sigma(\vec{r})^{2}\rangle \equiv
\langle(\sigma(\vec{r})-\langle\sigma(\vec{r})
\rangle)^{2} \rangle = c (1-c )(\sigma_{0}
-\sigma_{1})^{2}
\end{equation}

\noindent If we define the correlation on a lattice with
spacing of the domain size $\pi a^{2}$ as
$\langle\delta\sigma(\vec{r})\delta\sigma(\vec{r\,'})
\rangle \equiv \epsilon\,\delta(\vec{r}-\vec{r}\,')
\simeq\epsilon\,\delta_{ij}/\pi a^{2}$, we find,

\begin{equation}
\epsilon \simeq \pi a^{2} c
(1-c)(\sigma_{0}-\sigma_{1})^{2}.
\end{equation}

\noindent Using this result in (\ref{ALPHA}), we find that
the attenuation $\alpha$ for small area fractions $c$
compares with that found in Section III, equations
(\ref{SIGMA}) and (\ref{BEER}).

\section{Asymptotic Form of the Uniform Surface
Green's Function}

We briefly outline the calculation of the asymptotic forms
of the Green's function matrix for a nonviscous
compressible two dimensional fluid at the interface of an
incompressible viscous substrate fluid. The zeroes of the
determinant of the matrix ${\bf L}(\vec{q})$,
\end{multicols}
\widetext

\begin{eqnarray}
\left[(\omega^{2}-\sigma\vert q \vert q^{2}+2i\omega
\nu q^{2})({iB\over \omega} q^{2}l+2 \nu
q^{2}-i\omega)+ q^{2} (\sigma q^{2}-2i\omega \nu l)
({iB\over \omega}q^{2}+2\nu \vert q \vert )\right]
\left[ 1+{i\nu q^{2}\over \omega} \right]
\end{eqnarray}

\begin{multicols}{2}
\narrowtext

\noindent are shown in Fig. 8 and define the modes of the
system. Note that these eigenmodes are not orthogonal
and are hence coupled to each other \cite{EARN2}.  The
Green's function is found by Fourier transforming each
element of ${\bf L}^{-1}(\vec{q})$ via the contours  shown
in Figure 3.  Contributions from the shear mode near the
branch point at $q= \sqrt{i\omega /\nu}$ and the branch
cut at are neglected as well as the integrations on the
imaginary axes. The leading behavior of the Green's
function is,

\begin{equation}
G_{11}(r) \simeq {-\omega \over 6p\sigma_{0}}
H_{0}^{(1)}(k_{t}r)+{3\omega\nu^{2} \over 8B_{0}^{2}}
H_{0}^{(1)}(k_{l}r)
\end{equation}

\begin{eqnarray}
G_{i1}(r) \simeq {\sqrt{\nu\omega}\,e^{3\pi i/4} \over 6
\sigma_{0}} \left( {r_{i} \over r}\right)
H_{1}^{(1)}(k_{t}r)+\;\;\;\;\;\;\nonumber \\
\;\;\;\;\;\;\;\;\;\;\;{e^{ i\pi /8} \nu^{5/4}\omega^{1/4}
\over 4B_{0}^{5/2}} \left( {r_{i} \over r}\right)
H_{1}^{(1)}(k_{l}r)
\end{eqnarray}

\noindent and,

\begin{eqnarray}
G_{1i}(r) \simeq {e^{3\pi i/4}\sqrt{\nu\omega}\over
6B_{0}}\left({r_{i} \over
r}\right) H_{1}^{(1)}(k_{t}r) -\;\;\;\;\;\;\;\nonumber \\
\;\;\;\;\;\;\;\;\;{e^{i\pi /8}(2B_{0}-\sigma_{0})\nu^{5/4}
\omega^{1/4}\over 4B_{0}^{5/2}}\left({r_{i} \over r}\right)
H_{1}^{(1)}(k_{l}r)
\end{eqnarray}

\noindent where the coefficients of the Hankel functions
are expansions in small viscosity of the residues at the
poles $k_{t}$ and $k_{l}.$ Though not needed in this
study, similar expressions can be found for
$G_{23}(\vec{r})$ and $G_{jj}(\vec{r})$.

\begin{figure}
\caption{(a) Circular domains of higher density
regions of surfactant coverage. Images were taken with
a Brewster angle microscope. (b) Stripe phases in the
``coexistence'' region.}
\end{figure}

\begin{figure}
\caption{Arrangement for  scattering capillary
waves from circular surfactant domains. The
wedge on the left is a wavemaker excited at frequency
$\omega$. Surface parameters that surfactant domains
change, $\chi$, are surface tension $\sigma$, surface
compressional modulus, $B$, and interfacial viscosities,
$\eta^{(2)}$ and $\zeta^{(2)}$.}
\end{figure}

\begin{figure}
\caption{A schematic surface pressure - concentration
phase diagram of an amphiphillic monolayer
($\sigma_{w}$ is the surface tension of a clean air/water
interface). The negative of the slopes of the isotherm just
outside the plateau multiplied by their respective
densities, $\Gamma_{j}$, are the isothermal surface
compressional moduli, $B_{j}$, of the two phases near
equilibrium.}
\end{figure}

\begin{figure}
\caption{(a) Real and imaginary shifts in the transverse
wavevector, $\Delta'$ and $\Delta''$ plotted for
$\sigma_{0} = 72$ dyne/cm, $\nu = 0.01$ cm$^{2}$/s,
and $\omega \simeq 5184$ s$^{-1}$. (b) Relative damping
rates between the transverse and longitudinal modes;
the transverse mode is more heavily damped for
parameters above the $y=0$ axis, longitudinal waves
are damped quickly below $y=0$.}
\end{figure}

\begin{figure}
\caption{(a) Limit when longitudinal mode is damped
near each scatterer. The transverse waves are weakly
damped on interdomain length scales $l$. (b)
Longitudinal waves are long-ranged, transverse waves
are strongly damped; $Im\{k_{t}\} \ll l^{-1}$.}
\end{figure}

\begin{figure}
\caption{Schematic of a proposed experiment to detect
scattering from interfacial inhomogeneities at the
air/water interface}
\end{figure}

\begin{figure}
\caption{(a) Diagram representing the sum in (A1). (b)
Green's function averaged over frozen disorder.}
\end{figure}

\begin{figure}
\caption{Approximate location of poles in the complex
$q$-plane. The shear mode in our analyses has a large
imaginary component. The contours are the integrals
used to evaluate the Green's functions}
\end{figure}

\end{multicols}

\end{document}